\documentclass[aps,prd,reprint]{revtex4-1}
\usepackage{graphicx}
\usepackage{dcolumn}
\usepackage{natbib,aas_macros}
\usepackage{color}
\usepackage{amssymb}
\usepackage{amsthm}
\usepackage{mathrsfs}
\usepackage{amsmath}

\newcommand{\msun}{{\rm M_\odot}}

\begin{document}
\title{Multi-messenger tests of cosmic-ray acceleration\\
in radiatively inefficient accretion flows}

\author{Shigeo S. Kimura$^{1,2,3,4,5}$}
\author{Kohta Murase$^{1,2,3,6}$}
\author{Peter M{\'e}sz{\'a}ros$^{1,2,3}$}

\affiliation{$^1$Department of Physics, The Pennsylvania State University, University Park, Pennsylvania 16802, USA}
\affiliation{$^2$Department of Astronomy \& Astrophysics, The Pennsylvania State University, University Park, Pennsylvania 16802, USA}
\affiliation{$^3$Center for Particle and Gravitational Astrophysics, The Pennsylvania State University, University Park, Pennsylvania 16802, USA}
\affiliation{$^4$Frontier Research Institute for Interdisciplinary Sciences, Tohoku University, Sendai 980-8578, Japan}
\affiliation{$^5$Astronomical Institute, Tohoku University, Sendai 980-8578, Japan}
\affiliation{$^6$Yukawa Institute for Theoretical Physics, Kyoto, Kyoto 606-8502 Japan}

\date{\today}

\begin{abstract}
The cores of active galactic nuclei (AGNs) have been suggested as the sources of IceCube neutrinos, and recent numerical simulations have indicated that hot AGN coronae of Seyfert galaxies and radiatively inefficient accretion flows (RIAFs) of low-luminosity AGNs (LLAGNs) may be promising sites of ion acceleration. 
We present detailed studies on detection prospects of high-energy multi-messenger emissions from RIAFs in nearby LLAGNs. 
We construct a model of RIAFs that can reproduce the observational features of the current X-ray observations of nearby LLAGNs. 
We then calculate the high-energy particle emissions from nearby individual LLAGNs, including MeV gamma rays from thermal electrons, TeV--PeV neutrinos produced by non-thermal protons, and sub-GeV to sub-TeV gamma rays from proton-induced electromagnetic cascades.
We find that, although these are beyond the reach of current facilities, proposed future experiments such as e-ASTROGAM and IceCube-Gen2 should be able to detect the MeV gamma rays and the neutrinos, respectively, or else they can place meaningful constraints on the parameter space of the model. 
On the other hand, the detection of high-energy gamma rays due to the electromagnetic cascades will be challenging with the current and near-future experiments, such as {\it Fermi} and Cherenkov Telescope Array. 
In an accompanying paper, we demonstrate that LLAGNs can be a source of the diffuse soft gamma-ray and TeV--PeV neutrino backgrounds, whereas in the present paper, we focus on the prospects for multi-messenger tests which can be applied to reveal the nature of the high-energy neutrinos and photons from LLAGNs.
\end{abstract}

\maketitle

\section{Introduction}\label{sec:intro}

The IceCube Collaboration reported the detection of extraterrestrial neutrinos in 2013 \cite{ice13prl,2013Sci...342E...1I} and provided the details of the diffuse neutrino background intensity \cite{ice14,ice15,ice15a}. The origin of the astrophysical neutrino background has yet to be confirmed (see Ref.~\cite{2017PTEP.2017lA105A} for a review). 
Gamma-ray bursts (GRBs) were expected to be a promising neutrino source \cite{WB97a,MW01a,2003PhRvL..91g1102D,GHA04a,MN06b}. However, stacking analyses using the information of the observed GRBs detect no associated neutrinos, which puts an upper limit on the GRB contribution to the neutrino intensity of $\lesssim1$\% \cite{IceCube16b,IceCube17b}. Note that these analyses focus on the prompt emission from bright GRBs, while neutrinos produced in the afterglow phase \cite{WB00a,MN06a,Mur07a,KMM17b}, by low-luminosity GRBs and engine-driven supernovae~\cite{mur+06,SMM16a,ZMK18a,2019ApJ...872..110B,2018arXiv181210289Z}, 
or by failed GRBs (also known as choked jets; \cite{MW01a,2003PhRvD..68h3001R,2005PhRvL..95f1103A,HA08a,mi13,HKN18a,DT18a,KMB18a}) are not very constrained.

Blazars are also believed to be capable of strong neutrino emission \cite{1992A&A...260L...1M,1997ApJ...488..669H,ad01}. 
Recently, IceCube reported the detection of a high-energy neutrino coincident with a flaring activity of a blazar, TXS 0506+056 \cite{Aartsen2018blazar1}. 
Thanks to the ensuing multi-messenger followup campaign (see \cite{Aartsen:2016lmt}), the broad-band spectral energy distribution during the flaring period has been determined, which enables one to model the neutrino emission in detail \cite{2018ApJ...864...84K,2018ApJ...865..124M,2018arXiv180705113L,2019MNRAS.483L..12C,2019NatAs...3...88G}. 
The IceCube Collaboration also found a neutrino flare from this object during 2014 -- 2015, by re-analyzing their archival data \cite{Aartsen2018blazar2}. 
However, this neutrino flare is not accompanied by a corresponding GeV gamma-ray flaring activity \cite{Aartsen:2019gxs}, which challenges the theoretical modeling of the neutrino emission \cite{2018ApJ...865..124M,2018arXiv181205654R,2018arXiv181205939R,2018arXiv180900601W}. 
Note that the coincident detection and the archival neutrino flare do not, however, mean that the blazars are the dominant source of the diffuse neutrinos. The stacking analyses of the blazars detected by Fermi result in a non-detection \cite{Aartsen:2016lir,Neronov:2016ksj,2019arXiv190406371Y,2019JCAP...02..012H}, 
which implies that their contribution is less than $\sim10-30$ \% of the total astrophysical neutrinos. 
Also, the absence of event clustering in the arrival distribution of neutrinos indicates that the contributions from flaring blazars should be less than $\sim10-50$\%~\cite{Murase:2016gly,2018ApJ...865..124M}.

Another constraint is provided by the extragalactic gamma-ray background detected by {\it Fermi} \cite{Ackermann:2014usa}. When astrophysical neutrinos are produced through pion decay, gamma rays are also produced simultaneously. The generated gamma-ray luminosity is comparable to the neutrino luminosity, and the TeV--PeV gamma rays are cascaded down to the GeV--TeV energy range during their propagation towards Earth. In order to avoid overproducing the observed extragalactic gamma-ray background, the neutrino spectral index should be smaller than $2.1-2.2$ \cite{mal13}, which is in tension with the best-fit spectrum of the observed neutrinos in the shower analyses \cite{ice15a,ice15,Aartsen:2015rwa,Aartsen:2017kru}. 
Also, the neutrino flux at 1--100 TeV is higher than that above 100 TeV \cite{ice15,ice15a}, although this might be due to the strong atmospheric background \cite{2018A&A...615A.168P}. If such a ``medium-energy excess'' is real, the serious tension with the gamma-ray background is unavoidable, suggesting that the main sources are opaque and hidden in high-energy gamma rays \cite{Murase:2015xka}. 
This argument disfavors many astrophysical scenarios as the origin of these neutrinos, including starburst galaxies \cite{lw06,mal13,2013PhRvD..87f3011H,2014PhRvD..89l7304A,2014PhRvD..89h3004L,tam14,sen+15,2016ApJ...826..133X,2017ApJ...836...47B,2018PASJ...70...49S}, 
galaxy clusters \cite{min08,kot+09,mal13,2015A&A...578A..32Z,2016ApJ...828...37F,FM18a}, 
and radio-galaxies \cite{bec+14,2016JCAP...09..002H}. 

We consider high-energy neutrino emission from the vicinity of supermassive black holes (SMBHs) in active galactic nuclei (AGNs)~\cite{1981MNRAS.194....3B,ke86,1986ApJ...305...45Z,brs90,ste+91,1999MNRAS.302..373B,am04}. 
A luminous AGN hosts a geometrically thin, optically thick accretion disk that produces copious UV photons \cite{ss73,1983ApJ...268..582M,1987ApJ...321..305C}, and the ratio of the observed UV to X-ray luminosity is very high \cite{2004MNRAS.351..169M,2007ApJ...654..731H,2012MNRAS.425..623L}. Such target photon fields lead to a hard neutrino spectrum at PeV energies~\cite{2015JETP..120..541K,2019arXiv190400554I}. The accretion shock has been considered, but the existence of such a shock has not been supported by numerical simulations so far. On the other hand, recent studies on magnetorotational instabilities suggest that particle acceleration via magnetic reconnections and turbulence is promising in AGN coronae, and Ref.~\cite{2019arXiv190404226M} showed that the mysterious 10 -- 100~TeV component in the diffuse neutrino flux can be explained by the AGN core model of radio-quiet AGNs. It was found that the Bethe-Heitler process is critically important, which led to robust predictions of MeV gamma rays via proton-induced cascades. 

Low-luminosity AGNs (LLAGNs), however, have different spectral energy distributions, in which an UV bump is absent~\cite{ho08}. This indicates that there is an optically thin, hot accretion flow instead of an optically thick disk. Remarkably, plasma properties of hot AGN coronae and radiatively inefficient accretion flows (RIAF; \cite{ny94,YN14a}) in LLAGNs seem similar in the sense that the plasmas are expected to be collisionless for ions. It is natural to consider the same type of proton acceleration in both Seyfert galaxies and LLAGNs. 
Ref.~\cite{kmt15} considered the stochastic acceleration expected in such RIAFs of LLAGNs, and showed that the neutrinos produced by the accelerated protons can account for the diffuse astrophysical neutrino background (see also Refs. \cite{KG16a,2019MNRAS.483L.127R} for neutrino emissions from LLAGNs). The LLAGN model can avoid the gamma-ray and the point-source constraints, thanks to its compact emission region and high number density, although Ref.~\cite{kmt15} did not provide details of the resulting gamma-ray spectra. 


In this paper, we describe a refined LLAGN model, and show how multi-messenger information on neutrinos and gamma rays can be used as a test of the proposed LLAGN model. 
We estimate the physical quantities in the RIAFs of several nearby LLAGNs including the photons from the thermal electrons in Section \ref{sec:RIAF}.
We then estimate the high-energy proton spectra in Section \ref{sec:protons}, and calculate the high-energy neutrino spectra and their detectability in Section \ref{sec:neutrino}. 
We calculate the gamma rays from proton-induced electromagnetic cascades in Section \ref{sec:gamma}. 
Finally, we summarize the results and discuss their implications in Section \ref{sec:summary}. 
We note that our refined model can reproduce the diffuse MeV gamma-ray and the TeV -- PeV neutrino backgrounds simultaneously without overshooting the {\it Fermi} data, which is shown in an accompanying paper. In this paper, we focus on the detection prospects of individual nearby LLAGNs. 

\section{Physical quantities in RIAFs}\label{sec:RIAF}

We consider a RIAF of size $R$ and mass accretion rate $\dot M$ around a SMBH of mass $M_{\rm BH}$. We use the notation $Q_x=10^x$ in cgs units, unless otherwise noted. To represent the physical quantities in the RIAF, it is convenient to normalize $R$ by the Schwarzschild radius: $R=\mathcal R R_S\simeq 2.95\times10^{14}\mathcal R_1 M_8$, where $R_S=2GM_{\rm BH}/c^2$ is the Schwarzschild radius, $G$ is the gravitational constant, and $c$ is the speed of light. The mass accretion rate is  normalized by the Eddington accretion rate: $\dot m = \dot Mc^2/L_{\rm Edd}$, where $L_{\rm Edd}\simeq 1.3\times10^{46}M_8\rm~erg~s^{-1}$ is the Eddington luminosity.


According to recent magnetohydrodynamic (MHD) simulations (see e.g., Refs. \cite{MM03a,Mck06a,OM11a,2012MNRAS.426.3241N,2013MNRAS.428.2255P,2019MNRAS.485..163K}), the radial velocity, the sound velocity, the scale height, the number density, the magnetic field, and the Alfven velocity in the RIAF are estimated to be
\begin{eqnarray*}
& V_R&\approx\frac12 \alpha V_K \simeq 3.4\times10^8 \mathcal R_1^{-1/2} \alpha_{-1} \rm~cm~s^{-1} \\
& C_s&\approx\frac12 V_K \simeq 3.4\times10^9 \mathcal R_1^{-1/2} \rm~cm~s^{-1} \\
& H&\approx\frac12 R \simeq 1.5\times10^{14}\mathcal R_1M_8\rm~cm  \\
& n_p&\approx \frac{\dot M}{4\pi m_p R H V_R}\simeq 4.6\times10^8\mathcal R_1^{-3/2}\alpha_{-1}^{-1}M_8^{-1} \dot m_{-2} \rm~cm^{-3}  \\
& B&\approx \sqrt{\frac{8\pi P_g}{\beta}}\simeq 2.6\times10^2 \mathcal R_1^{-5/4}\alpha_{-1}^{-1/2}M_8^{-1/2}\dot m_{-2}^{1/2}\beta_{0.5}^{-1/2} \rm~G \\
& V_A&\approx \frac{B}{\sqrt{4\pi m_p n_p}}\simeq 2.7\times10^9\mathcal R_1^{-1/2}\beta_{0.5}^{-1/2} \rm~cm~s^{-1} 
\end{eqnarray*}
where $V_K=\sqrt{GM_{\rm BH}/R}$ is the Keplerian velocity, $\alpha$ is the viscous parameter \cite{ss73}, $m_p$ is the proton mass, $\beta=8\pi P_g/B^2$ is the plasma beta, and $P_g=m_pn_pC_s^2$ is the gas pressure. We assume pure proton composition for simplicity. 
The magnetic field strength in the hot accretion flows depends on the configuration of the magnetic field: $\beta\sim 10-100$ for standard and normal evolution (SANE) flows, whereas $\beta\sim 1-10$ for magnetically arrested disks (e.g., \cite{Mck06a,2012MNRAS.426.3241N,2017MNRAS.467.3604R,2018MNRAS.478.5209C}). We use $\beta\sim 3.2$ as a reference value because lower $\beta$ plasma are suitable for producing non-thermal particles \cite{2018ApJ...862...80B}. For the viscous parameter $\alpha$, SANE models tend to give a lower value, $\alpha\simeq 0.03$ \cite{2013MNRAS.428.2255P,2019MNRAS.485..163K}, while observations of X-ray binaries and dwarf novae suggest $\alpha\simeq0.1-1$ (see Ref. \cite{2019arXiv190101580M} and references therein). Here, we set $\alpha=0.1$ as a reference value.


Although cooling processes have little influence on the dynamical structure in the RIAF, the thermal electrons supply target photons for photohadronic interactions and $\gamma\gamma$ two-photon annihilation. We calculate the characteristics of the target photons in the RIAF using a method similar to Ref.~\cite{kmt15}. 
We consider synchrotron, bremsstrahlung, and inverse Compton emission processes. The calculation method of the emission spectrum due to each process was discussed in the Appendix of Ref.~\cite{kmt15}. Note that this treatment is valid only for flows with Thomson optical depths $\tau_T \approx n_p \sigma_T R < 1$, where $\sigma_T$ is the Thomson cross section.

As long as $\dot m \gtrsim 10^{-2}\alpha^2\sim10^{-4}\alpha_{-1}^2$, the balance between the cooling rate and heating rate of the thermal electrons determines the electron temperature, $\Theta_e = k_B T_e/(m_e c^2)$, where $m_e$ is the electron mass and $k_B$ is the Boltzmann constant \cite{xy12,YN14a}. Then, the electron heating rate is equal to the bolometric luminosity from the thermal electrons.  If the Coulomb collisions with the thermal protons are the dominant heating process, the heating rate is proportional to $n_p^2$, which leads to $L_{\rm bol}\propto \dot m^2$. Then, the bolometric luminosity is phenomenologically given by (see, e.g., Ref. \cite{mq97,KFM08a})
\begin{equation}
L_{\rm bol}\approx \epsilon_{\rm rad,sd}\left(\frac{\dot m}{\dot m_{\rm crit}}\right)^2 \dot m_{\rm crit}L_{\rm Edd},\label{eq:Lbolmdot}
\end{equation}
where $\dot m_{\rm crit}$ is the normalized critical accretion rate above which the RIAF solution no longer exists \cite{ny95,ACK95a} and $\epsilon_{\rm rad,sd}\sim0.1$ is the radiation efficiency of the standard thin disk. The critical accretion rate can be expressed as a function of $\alpha$ \cite{1997ApJ...477..585M,xy12}. Following Ref. \cite{1997ApJ...477..585M}, we represent $\dot m_{\rm crit}\sim 3\alpha^2\simeq 3\times10^{-2}\alpha_{-1}^2$. Note that the dissipation processes in collisionless accretion flows are still controversial. If the electrons are directly heated by plasma dissipation processes induced by kinetic instabilities \cite{qg99,sha+07,riq+12,KSS14a,hos13,hos15,2015ApJ...800...88S,2015ApJ...800...89S,2018arXiv180901966Z}, the electron heating rate may be proportional to $\dot m$, leading to $L_{\rm bol}\propto \dot m$ as assumed in Ref. \cite{kmt15}. In reality, the scaling relation may be located between the two regimes. In this paper, we use Equation (\ref{eq:Lbolmdot}) for simplicity.

Observations give us the X-ray luminosity, $L_X$, which is connected to $\dot m$ in our model. Using the bolometric correction factor, $\kappa_{{\rm bol}/X}$, the X-ray luminosity is related to the bolometric luminosity as 
\begin{equation}
L_{\rm bol}\approx \kappa_{{\rm bol}/X} L_X \label{eq:LbolLx}
\end{equation}
According to the X-ray surveys, $\kappa_{{\rm bol}/X}$ is higher for a higher $L_{\rm bol}$ or $\lambda_{\rm Edd}$, where $\lambda_{\rm Edd}=L_{\rm bol}/L_{\rm Edd}$ is the Eddington ratio. At the low-luminosity end, $\kappa_{{\rm bol}/X}$ becomes almost constant, $\kappa_{{\rm bol}/X}\sim 5-20$ \cite{2004ApJ...607L.107W,2012MNRAS.425..623L,2016MNRAS.459.1602L}. Using Equations (\ref{eq:Lbolmdot}) and (\ref{eq:LbolLx}) with a constant $\kappa_{{\rm bol}/X}$, we can write $\dot m$ as a function of observables:
\begin{equation}
 \dot m \approx  \left(\frac{\kappa_{{\rm bol}/X}L_X\dot m_{\rm crit}}{\epsilon_{\rm rad,sd}L_{\rm Edd}}\right)^{1/2}\approx 1.9\times10^{-2}M_8^{-1/2}L_{X,42}^{1/2}\alpha_{-1}\label{eq:mdot},
\end{equation}
where we use $\kappa_{{\rm bol}/X}=15$ and $\epsilon_{\rm rad,sd}=0.1$. This $\dot m$ is less than $\dot m_{\rm crit}$. Hence, typical LLAGNs with $L_X\lesssim 10^{42}\rm~erg~s^{-1}$ can host RIAFs.

\begin{figure}[tb]
   \begin{center}
    \includegraphics[width=\linewidth]{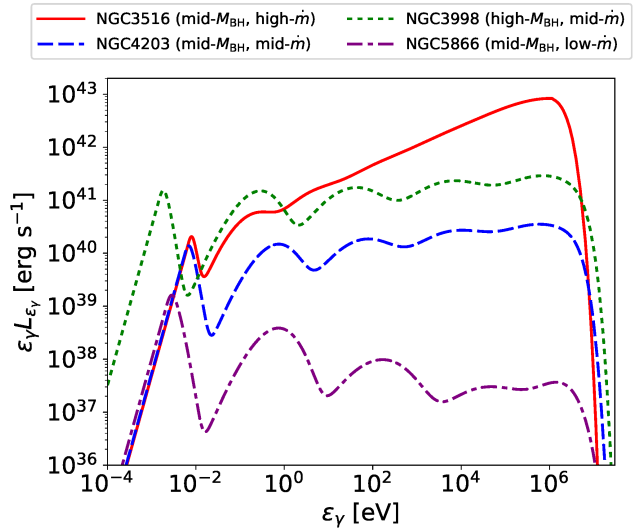}
    \caption{Soft photon spectra for NGC 3516 (red-solid line), NGC 4203 (blue-dashed line), NGC 3998 (green-dotted line), and NGC 5866 (purple-dot-dashed). }
    \label{fig:soft}
   \end{center}
  \end{figure}

We calculate spectral energy distributions of nearby LLAGNs listed in Table A.3 of Ref. \cite{2018A&A...616A.152S}, which provides $M_{\rm BH}$, $L_X$, luminosity distance ($d_L$), and declination angle ($\delta$) for 70 LLAGNs. The mass accretion rate of the listed LLAGNs is estimated using Equation (\ref{eq:mdot}) with $\kappa_{{\rm bol}/X}=15$. 
We find that 7 of them have standard disks, i.e., $\dot m > \dot m_{\rm crit}$, while the others host RIAFs. Figure \ref{fig:soft} shows the target photon spectra from 4 LLAGNs whose parameters and resulting physical quantities are tabulated in Table \ref{tab:param} and \ref{tab:quant}, respectively.  The values of the other parameters are tabulated in Table \ref{tab:fixed}. The four LLAGNs differ in $M_{\rm BH}$ and $\dot m$. NGC 3516, NGC 4203, and NGC 5866 have $M_{\rm BH}$ close to $10^8~\msun$, while NGC 3998 hosts a SMBH of $M_{\rm BH}\sim10^9~\msun$. $\dot m$ is close to the critical accretion rate for NGC 3516, $\dot m\sim 0.1\dot m_{\rm crit}$ for NGC 4203 and NGC 3998, and $\dot m\sim 0.01 \dot m_{\rm crit} $ for NGC 5866. 
For all the LLAGNs, the synchrotron emission peaks in the radio band. For LLAGNs with $\dot m \gtrsim 10^{-2}\dot m_{\rm crit}$, the inverse Compton emission of the synchrotron photons produces infrared to MeV photons. For lower $\dot m$ cases, the bremsstrahlung emits MeV photons due to inefficient Comptonization. The inverse Compton emission spectrum is hard and smooth for higher $\dot m$, while it is soft and bumpy for lower $\dot m$ due to a high value of electron temperature and a low value of Compton-$Y$ parameter, $y\approx \tau_T(4 \Theta_e + 16\Theta_e^2)$ (see Table \ref{tab:param} for the values of $\dot m$, $\Theta_e$, and $\tau_T$). A high value of $M_{\rm BH}$ with fixed $\dot m$ lowers the peak frequency of the synchrotron emission due to the weak magnetic field, and increases the entire luminosity because of a high net accretion luminosity, $\dot Mc^2 = \dot m L_{\rm Edd}\propto M_{\rm BH}$.

\begin{figure}[tb]
   \begin{center}
    \includegraphics[width=\linewidth]{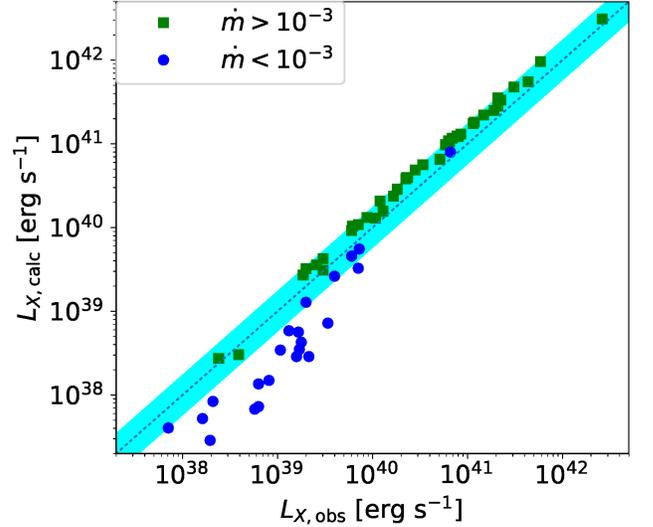}
    \caption{Relationship between the observed X-ray luminosity, $L_{X,\rm obs}$, and the X-ray luminosity obtained by the model calculation, $L_{X,\rm calc}$. The green squares are LLAGNs with $\dot m>10^{-3}$, while the blue circles are those with $\dot m < 10^{-3}$. The dotted line represents $L_{X,\rm obs}=L_{X,\rm calc}$, and cyan band indicates $L_{X,\rm obs}/1.7 < L_{X,\rm calc}< 1.7 L_{X,\rm obs}$, in which all the green squares are located.   }
    \label{fig:Lx}
   \end{center}
  \end{figure}

Next, we compare the X-ray luminosities obtained by our calculations and observations. Figure \ref{fig:Lx} shows the relation between the observed $2-10$ keV X-ray luminosity, $L_{X,\rm obs}$, and the X-ray luminosity calculated by our model, $L_{X,\rm calc}$ in the same band. Intriguingly, our simple model is in a good agreement with the observations for $\dot m > 10^{-3}$. The two luminosities match within a factor of 1.7 in this sample. We stress that we do not adjust the X-ray luminosity but we calculate photon spectra with the one-zone model using $\dot m$ estimated by Equations (\ref{eq:LbolLx}) and (\ref{eq:mdot}). 
For a lower value of $\dot m < 10^{-3}$, the synchrotron emission is more efficient than the inverse Compton emission. This causes a higher value of $\kappa_{{\rm bol}/X}$, resulting in $L_{X,\rm calc}<L_{X,\rm obs}$ as seen in the figure. For nearby low-ionization nuclear emission-like regions (LINERs), the bolometric correction factor is estimated to be $\kappa_{{\rm bol}/X}\sim 50$ \cite{ehf10}.  $L_{X,\rm calc}$ is higher with such a higher value of $\kappa_{{\rm bol}/X}$, since it leads to a higher value of $\dot m$. Hence, a higher value of $\kappa_{{\rm bol}/X}$ is more consistent with our model with $\dot m<10^{-3}$. Nevertheless, we use $\kappa_{{\rm bol}/X}=15$ because LLAGNs with $\dot m < 10^{-3}$ do not affect the detectability of high-energy neutrinos as shown in Section \ref{sec:neutrino}.  

The bright LLAGNs are detected by the {\it Swift} BAT,  most of which show hard X-ray spectra. 
Thus, very interestingly, our model is consistent with the BAT data in terms of luminosity. 
In addition, RIAF models generally predict that a higher $\dot m$ object has a harder photon spectrum in the X-ray band owing to a higher value of the Compton-$Y$ parameter, which is consistent with the observed anti-correlation between the X-ray spectral index and the Eddington ratio \cite{2009MNRAS.399..349G,2018ApJ...859..152S}.

Recently, Ref.~\cite{2019ApJ...870...73Y} estimated the cutoff energy in X-ray spectrum in NGC 3998 to be around 100 keV using the NuSTAR and XMM-Newton data. However, they just measured a slight softening of the spectrum, which can be reconciled by our RIAF model. The Compton scattering makes a few bumps in the broad-band spectrum, which causes a softening in the X-ray band for NGC 3998 as seen in Figure \ref{fig:soft}. Here, we do not compare our model to observations in detail, because they are beyond the scope of this paper.

\begin{figure*}[tb]
   \begin{center}
    \includegraphics[width=\linewidth]{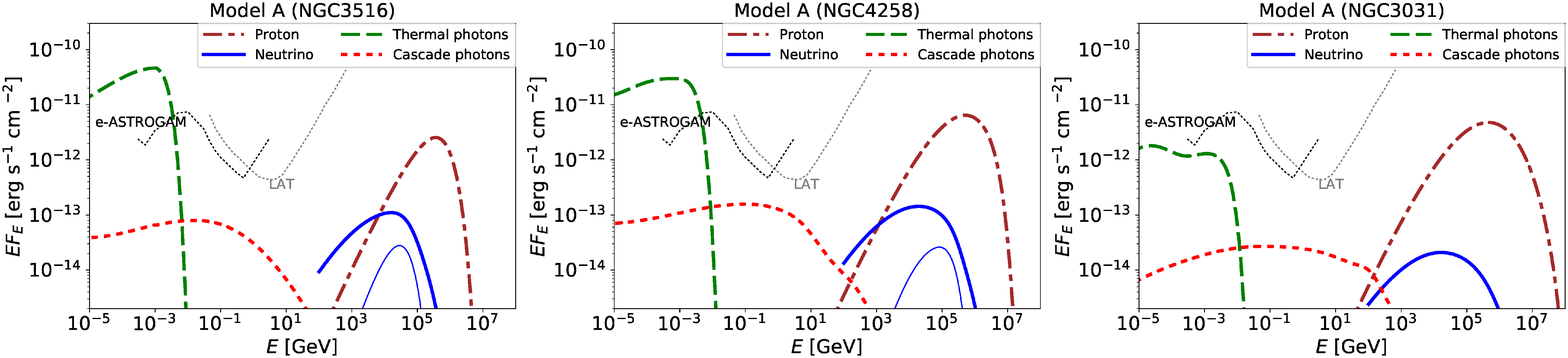}
    \includegraphics[width=\linewidth]{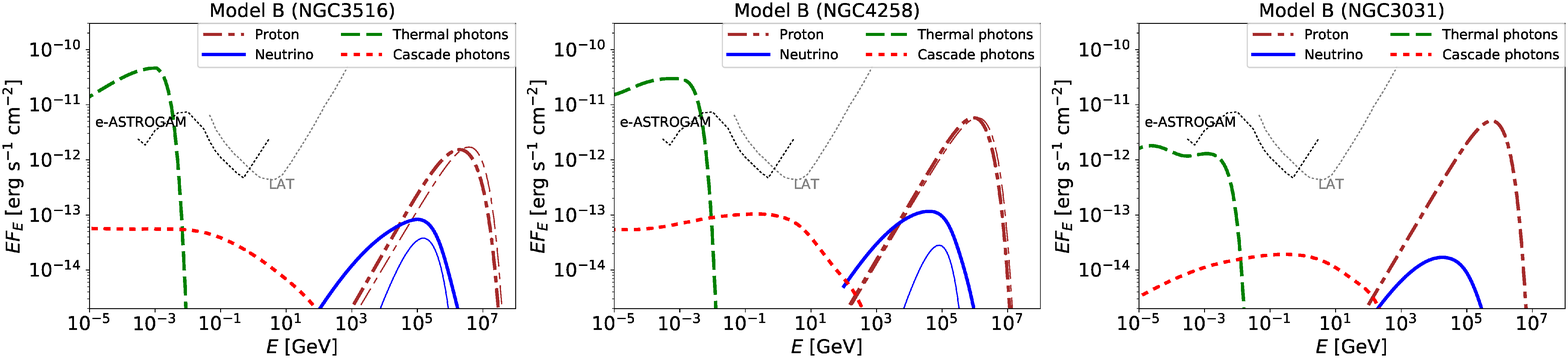}
    \includegraphics[width=\linewidth]{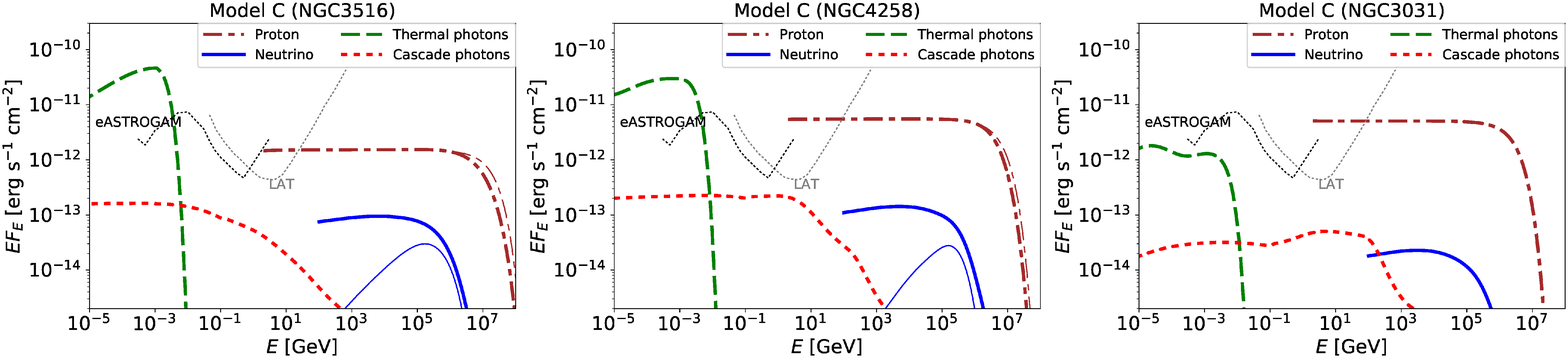}
    \caption{Spectral energy distributions of gamma-ray (dashed by thermal electrons; dotted by hadronic cascade), neutrino (solid), and proton (dot-dashed) fluxes for Model A (stochastic acceleration; left panel), Model B (power-law injection with $s_{\rm inj}=1.0$; middle), and Model C (power-law injection with $s_{\rm inj}=2.0$; right), respectively. 
The upper, middle, and lower panels are for NGC 3516, NGC 4258, and NGC 3031, respectively. The thin-dot-dashed lines in the middle and right columns depict the injection spectrum of the protons. The thin-solid lines are the $p\gamma$ neutrino flux. For NGC 3031, the $p\gamma$ neutrino flux is below the lower end of the figure.}
    \label{fig:ELE}
   \end{center}
  \end{figure*}

In order to obtain the electron temperature more concretely, we need to detect a clear cutoff feature above 100 keV. We plot the photon spectra due to thermal electrons above 10 keV with the sensitivity curve of the proposed future satellite, e-ASTROGAM \cite{2017ExA....44...25D} in Figure \ref{fig:ELE}. The MeV gamma rays will be easily detected for NGC 3516 and NGC 4258, although it is not expected for NGC 3031. The other proposed MeV gamma-ray satellites, AMEGO \cite{2017ICRC...35..798M} and GRAMS \cite{2020APh...114..107A}, have similar or better sensitivity in this range. The MeV observations of nearby LLAGNs will provide not only the electron temperature in RIAFs for the first time, but also the crucial test for the LLAGN contribution to the MeV gamma-ray background (see the accompanying paper).

\begin{table}[tb]
 \begin{center}
\caption{Observational quantities for nearby LLAGNs. These LLAGNs are selected as the ten brightest ones in X-ray band except NGC 5866, which is an LLAGN with a lower accretion rate shown in Figure \ref{fig:soft}. Units are [erg s$^{-1}$ cm$^{-2}$] for $F_{X,\rm obs}$,  [erg s$^{-1}$] for $L_{X,\rm obs}$, [$\rm M_{\odot}$] for $M_{\rm BH}$, [Mpc] for $d_L$, and [deg] for $\delta$.  \label{tab:param}}
\begin{tabular}{c|cccccc}
\hline
 ID  & Type & $\log F_{X,\rm obs}$ & $\log L_{X,\rm obs}$ & $\log M_{\rm BH}$ & $d_L$ & $\delta$ \\
NGC & & [erg s$^{-1}$ cm$^{-2}$] & [erg s$^{-1}$] & [$\msun$] & [Mpc] & [deg]  \\
\hline
4565 & S1.9 & -10.73 & 41.32 & 7.43 & 9.7 & 26.0 \\
3516 & S1.2 & -10.84 & 42.42 & 8.07 & 38.9 & 72.6 \\
4258 & S1.9 & -10.84 & 40.90 & 7.62 & 6.8 & 47.3 \\
3227 & S1.5 & -11.06 & 41.64 & 7.43 & 20.6 & 19.9 \\
4138 & S1.9 & -11.26 & 41.28 & 7.17 & 17.0 & 43.7 \\
3169 & L2 & -11.32 & 41.35 & 8.16 & 19.7 & 3.5 \\
4579 & S1.9/L & -11.36 & 41.17 & 7.86 & 16.8 & 11.8 \\
3998 & L1.2 & -11.43 & 41.32 & 9.23 & 21.6 & 55.5 \\
3718 & L1.9 & -11.48 & 41.06 & 7.77 & 17.0 & 53.1 \\
4203 & L1.9 & -11.68 & 40.37 & 7.89 & 9.7 & 33.2 \\
4486 & L2 & -11.71 & 40.82 & 9.42 & 16.8 & 12.4 \\
3031 & S1.5 & -11.71 & 39.48 & 7.82 & 3.6 & 69.1 \\
\hline
5866 & T2 & -14.16 & 38.29 & 7.92 & 15.3 & 55.8  \\
\hline
\end{tabular}
 \end{center}
\end{table}

\begin{table*}[tb]
 \begin{center}
\caption{Physical quantities of the RIAF in the nearby LLAGNs. The values of $L_p$ and $P_{\rm CR}/P_g$ are for models A/B/C. Units are [cm] for $R$, [cm$^{-3}$] for $n_p$, [G] for $B$, [MeV] for $\varepsilon_{\gamma\gamma}$, and [erg s$^{-1}$] for $L_p$. \label{tab:quant}}
\begin{tabular}{|c||c|c|c|c|c|c|c|c|c|}
\hline
 ID  & $\log\dot m$ & $\log R$ & $\log n_p$ & $\log B$ & $\log \tau_T$ & $\theta_e$ & $\log \varepsilon_{\gamma\gamma}$ & $\log L_p$ & $P_{\rm CR}/P_g$  \\
NGC  & & [cm] & [cm$^{-3}$] & [G] & & &  [MeV] &  [erg s$^{-1}$]  & [\%] \\
\hline
4565 & -1.78 & 13.90 & 9.45 & 2.81 & -0.83 & 1.09 & 2.78 & 41.23/41.05/41.74 & 10/6/37 \\
3516 & -1.55 & 14.54 & 9.04 & 2.61 & -0.60 & 0.93 & 2.22 & 42.10/41.92/42.61 & 8/4/29 \\
4258 & -2.08 & 14.09 & 8.96 & 2.57 & -1.13 & 1.39 & 3.50 & 41.11/40.94/41.63 & 12/8/44 \\
3227 & -1.62 & 13.90 & 9.61 & 2.89 & -0.67 & 0.96 & 2.39 & 41.39/41.21/41.90 & 9/5/32 \\
4138 & -1.67 & 13.64 & 9.82 & 3.00 & -0.72 & 0.99 & 2.51 & 41.08/40.90/41.59 & 9/6/34 \\
3169 & -2.13 & 14.63 & 8.37 & 2.27 & -1.18 & 1.47 & 3.63 & 41.61/41.43/42.13 & 12/8/44 \\
4579 & -2.07 & 14.33 & 8.73 & 2.45 & -1.12 & 1.39 & 3.48 & 41.37/41.19/41.89 & 12/8/43 \\
3998 & -2.68 & 15.70 & 6.75 & 1.46 & -1.73 & 2.25 & 4.52 & 42.13/41.95/42.65 & 14/10/50 \\
3718 & -2.08 & 14.24 & 8.81 & 2.49 & -1.13 & 1.39 & 3.50 & 41.27/41.09/41.79 & 12/8/43 \\
4203 & -2.48 & 14.36 & 8.29 & 2.23 & -1.53 & 1.84 & 4.12 & 40.98/40.81/41.51 & 14/9/49 \\
4486 & -3.02 & 15.89 & 6.22 & 1.20 & -2.07 & 2.74 & 5.56 & 41.97/41.80/42.50 & 15/10/52 \\
3031 & -2.89 & 14.29 & 7.95 & 2.06 & -1.94 & 2.30 & 5.14 & 40.50/40.33/41.03 & 15/10/52 \\
\hline
5866 & -3.54 & 14.39 & 7.20 & 1.69 & -2.59 & 2.85 & 5.89 & 39.96/39.82/40.58 & 16/12/66 \\
\hline
\end{tabular}
 \end{center}
\end{table*}

\begin{table}[tb]
\begin{center}
\caption{Parameters in our models. \label{tab:fixed}}
 Common parameters
 
\begin{tabular}{ccccccc}
\hline
$\alpha$ & $\beta$ & $\mathcal R$  & $\kappa_{{\rm bol}/X}$ & $\epsilon_{\rm rad,sd}$ \\
\hline
0.1& 3.2 & 10 & 15 & 0.1  \\
\hline
\end{tabular}
 
Model-dependent parameters and quantities
\begin{tabular}{cccccccc}
\hline
Parameters &  $\epsilon_p$& $\zeta$ & $q$ & $s_{\rm inj}$  &  $\eta_{\rm acc}$ \\
\hline
Model A & 3.0$\times10^{-3}$ & 7.5$\times10^{-3}$ & 1.666 & - & - \\
Model B & 2.0$\times10^{-3}$ & - & - & 1.0 & $1.0\times10^6$ \\
Model C & 0.010 & - & - & 2.0 & $2.0\times10^5$ \\
\hline
\end{tabular}
\end{center}
\end{table}

\section{Non-thermal Protons in RIAFs}\label{sec:protons}

High-energy protons may be accelerated and injected into RIAFs by magnetic reconnections \cite{2001ApJ...562L..63Z,KGL12a,2014ApJ...783L..21S,hos15,KSQ16a,2018ApJ...862...80B}, stochastic acceleration via MHD turbulence \cite{KTS16a,2018PhRvL.121y5101C,2019MNRAS.485..163K,2019arXiv190103439W}, or electric potential gaps in the black hole magnetosphere \cite{2018ApJ...863L..31C,2018A&A...616A.184L}. 
We examine three cases of non-thermal proton spectra. 
One is the stochastic acceleration model (model A), in which we solve the diffusion equation in momentum space. 
The others are the power-law injection models (models B and C) in which we consider an injection term with a single power-law with an exponential cutoff. Such a power-law model mimics a generic acceleration process.

\subsection{Plasma condition}

For stochastic acceleration via turbulence to work, the relaxation time in the RIAF needs to be longer than the dissipation time, i.e., the plasma is collisionless. The relaxation time due to Coulomb collisions is estimated to be (e.g. Refs.~\cite{tk85,ktt14})
\begin{eqnarray}
t_{\rm rlx}\approx \frac{4\sqrt\pi}{\ln\Lambda n_p \sigma_Tc}\left(\frac{m_p}{m_e}\right)^2\left(\frac{C_s}{c}\right)^3\\
\simeq 1.8\times10^8 \alpha_{-1} M_8 \dot m_{-2}^{-1}\rm~s,
\end{eqnarray}
where $\ln\Lambda\sim20$ is the Coulomb logarithm. Interestingly, the relaxation time is independent of the normalized radius, $\mathcal R$.
The dissipation time in the accretion flow is represented as $t_{\rm diss}\sim \alpha^{-1}R/V_K$ \cite{pri81,KFM08a}. In the RIAF, this timescale is of the order of the infall time:
\begin{equation}
t_{\rm fall}\approx \frac{R}{V_R} \simeq 8.8\times10^5 \mathcal R_1^{3/2} \alpha_{-1}^{-1}M_8 \rm~s.\label{eq:tfall}
\end{equation}
Equating these two timescales, we obtain the critical radius within which the flow becomes collisionless (see also Ref. \cite{mq97}):
\begin{equation}
\mathcal R_{\rm crit}\simeq 35 \alpha_{-1}^{4/3}\dot m_{-2}^{-2/3}. 
\end{equation}
As long as $\dot m\lesssim \dot m_{\rm crit}$ with a fixed value of $\alpha\gtrsim0.1$, the RIAF consists of collisionless plasma at $R\lesssim 10 R_S$. 
Hence, one may naturally expect non-thermal particle production there. On the other hand, another accretion regime with a higher luminosity, such as the standard disk \cite{ss73} and the slim disk \cite{abr+88}, are made up of collisional plasma because the density and temperature there are orders of magnitude higher and lower than that in the RIAF, respectively. 
Therefore, particle acceleration is not guaranteed due to the thermalization via Coulomb collisions. 

\subsection{Stochastic acceleration model (A)}

In the stochastic acceleration model, protons are accelerated through scatterings with the MHD turbulence. The proton spectrum is obtained by solving the diffusion equation in momentum space (e.g., Refs. \cite{BE87a,2012SSRv..173..535P}):
\begin{equation}
\frac{\partial \mathcal F_p}{\partial t} = \frac{1}{\varepsilon_p^2}\frac{\partial}{\partial \varepsilon_p}\left(\varepsilon_p^2D_{\varepsilon_p}\frac{\partial \mathcal F_p}{\partial \varepsilon_p} + \frac{\varepsilon_p^3}{t_{\rm cool}}\mathcal F_p\right) -\frac{\mathcal F_p}{t_{\rm esc}}+\dot {\mathcal F}_{p,\rm inj},\label{eq:FP}
\end{equation}
where $\mathcal F_p$ is the momentum distribution function ($dN/d\varepsilon_p = 4\pi p^2 \mathcal F_p/c$), $D_{\varepsilon_p}$ is the diffusion coefficient, $t_{\rm cool}$ is the cooling time, $t_{\rm esc}$ is the escape time, and $\dot {\mathcal F}_{p,\rm inj}$ is the injection term to the stochastic acceleration. Considering resonant scatterings with Alfven waves, the diffusion coefficient is represented as \cite{dml96,sp08,2016ApJ...816...24K}
\begin{equation}
D_{\varepsilon_p}\approx\frac{\zeta c}{H}\left(\frac{V_A}{c}\right)^2\left(\frac{r_L}{H}\right)^{q-2}\varepsilon_p^2,
\end{equation}
where $r_L=\varepsilon_p/(eB)$ is the Larmor radius, $\zeta\approx8\pi\int P_k dk/B^2$ is the turbulent strength parameter, and $q$ is the power-law index of the turbulence power spectrum. The acceleration time is given by $t_{\rm acc}\approx \varepsilon_p^2/D_{\varepsilon_p}$.
We use a delta-function injection: $\dot {\mathcal F}_{p,\rm inj}=\dot {\mathcal F}_0 \delta(\varepsilon_p - \varepsilon_{\rm inj})$, where $\dot {\mathcal F}_0$ is normalization factor. We normalize the luminosity of the non-thermal protons so that the proton luminosity is a constant fraction of the accretion luminosity:
\begin{equation}
\int L_{\varepsilon_p}d\varepsilon_p = \epsilon_p \dot m L_{\rm Edd},
\end{equation}
where $L_{\varepsilon_p} = \varepsilon_p  t_{\rm loss}^{-1}dN/d\varepsilon_p$ is the differential proton luminosity ($t_{\rm loss}^{-1}=t_{\rm cool}^{-1}+t_{\rm esc}^{-1}$ is the total loss rate) and $\epsilon_p$ is the non-thermal proton production efficiency. We use the Chang \& Cooper method to solve the equation \cite{cc70,1996ApJS..103..255P}, and calculate the time evolution until steady state is achieved. Note that the normalization is different from that used in Ref. \cite{2019arXiv190404226M}, where we normalized the injection such that $\dot {\mathcal F}_0=f_{\rm inj}L_{X,\rm obs}/(4\pi^2\varepsilon_{\rm inj}^3R^3)$. Here, $f_{\rm inj}$ is the efficiency of the injection to the stochastic acceleration, and $f_{\rm inj}$ needs to be much smaller than $\epsilon_p$.

\subsection{Power-law injection models (B and C)}
For models B and C, we consider a generic acceleration mechanism, and the steady-state proton spectrum, $N_{\varepsilon_p}=dN/d\varepsilon_p$, is obtained by solving the transport equation: 
\begin{equation}
\frac{d}{d\varepsilon_p}\left(-\frac{\varepsilon_p}{t_{\rm cool}}N_{\varepsilon_p}\right) = \dot N_{\varepsilon_p,\rm inj} - \frac{N_{\varepsilon_p}}{t_{\rm esc}},
\end{equation}
where $\dot N_{\varepsilon_p,\rm inj}$ is the injection function. We consider a power-law injection with an exponential cutoff: 
\begin{equation}
\dot N_{\varepsilon_p,\rm inj} = \dot N_0 \left(\frac{\varepsilon_p}{\varepsilon_{p,\rm cut}}\right)^{-s_{\rm inj}}\exp\left(-\frac{\varepsilon_p}{\varepsilon_{p,\rm cut}}\right),\label{eq:dotNinj}
\end{equation}
where $\dot N_0$ is the normalization factor, $s_{\rm inj}$ is the injection spectral index, and $\varepsilon_{p,\rm cut}$ is the cutoff energy. We normalize the injection by 
\begin{equation}
\int \varepsilon_p \dot N_{\varepsilon_p,\rm inj}d\varepsilon_p = \epsilon_p \dot m L_{\rm Edd}.
\end{equation}
We can get an analytic solution of the transport equation (cf., Ref. \cite{2009herb.book.....D}):
\begin{equation}
N_{\varepsilon_p} = \frac{t_{\rm cool}}{\varepsilon_p}\int_{\varepsilon_p}^{\infty}d\varepsilon_p' \dot N_{p,\rm inj}(\varepsilon_p')\exp\left(-\mathcal G(\varepsilon_p,~\varepsilon_p')\right),
\end{equation}
\begin{equation}
 \mathcal G(\varepsilon_1,~\varepsilon_2) = \int_{\varepsilon_1}^{\varepsilon_2} \frac{t_{\rm cool}}{t_{\rm esc}}\frac{d\varepsilon_p'}{\varepsilon_p'}.
\end{equation}
This solution includes exponential term, so we  need to carefully treat the numerical integration.  In the rest of this paper, we show the results using Simpson's rule and 115 grid points per energy decade. We computed the numerical integration with the trapezoidal rule and/or with 50-200 grid points per decade, and confirmed that the error is reduced to less than 30\% using Simpson's rule with 100 grid points per energy decade.

The maximum achievable energy of protons is determined by the balance between acceleration and loss. We phenomenologically write the acceleration time as
\begin{eqnarray}
t_{\rm acc}\approx \eta_{\rm acc} \frac{r_L}{c}
\end{eqnarray}
where $\eta_{\rm acc}$ is a parameter for the acceleration timescale. Since the infall is the most efficient loss process for majority of the LLAGNs, we estimate the cutoff energy by $t_{\rm acc}=t_{\rm fall}$. This treatment approximates the cutoff energy within an error of a factor of a few.

\subsection{Escape and cooling timescales}\label{sec:loss}
High-energy protons escape from the RIAF via advection or diffusion.  The advective escape time is equal to the infall time given by Equation (\ref{eq:tfall}). The diffusive escape time depends on the magnetic field configuration. According to MHD simulations, the magnetic fields in RIAFs are stretched to the azimuthal direction. The non-thermal protons' mean free path perpendicular to the magnetic field is much shorter than that along the field line (e.g., Refs. \cite{KTS16a,2019MNRAS.485..163K}). In the turbulence with a power spectrum of $P_k \propto k^{-q}$, the parallel mean free path and the perpendicular diffusion coefficient  are estimated to be (e.g., Refs. \cite{1999ApJ...520..204G,CLP02a,sp08,2016ApJ...816...24K})
\begin{eqnarray}
 \lambda_\parallel \approx \frac{r_L}{3\zeta}\left(\frac{H}{r_L}\right)^{q-1},\\
D_\perp \approx \frac{D_\parallel}{1 + \left(\lambda_\parallel/r_L\right)^2}.
\end{eqnarray}
The Larmor radius in the RIAF is estimated to be
\begin{equation}
r_L\simeq 1.3\times10^{10}\varepsilon_{p,15}\mathcal R_1^{-5/4}\alpha_{-1}^{-1/2}M_8^{-1/2}\dot m_{-2}^{1/2}\beta_{0.5}^{-1/2} \rm~cm  ,
\end{equation}
with our fiducial parameter set (see Table \ref{tab:fixed}) and  $\varepsilon_{p,15}=\varepsilon_p/\rm PeV$.
Then, we obtain $\lambda_\parallel/r_L\simeq2.3\times10^4$, leading to $D_\perp/D_\parallel\simeq1.9\times10^{-9}$. Hence, we ignore the diffusive escape process in this paper, i.e., we use $t_{\rm esc}=t_{\rm fall}$. 
The value of $D_\perp$ could be larger due to possible cross-field diffusion. To understand the behavior of high-energy protons in configuration space, much more elaborate calculations would be required, which are beyond the scope of this paper (see Ref. \cite{2019MNRAS.485..163K} for related discussion).

As the proton cooling processes, we take into account $pp$ inelastic collisions, photomeson production, proton synchrotron processes, and the Bethe-Heitler process. The $pp$ cooling rate is 
\begin{equation}
t_{\rm pp}^{-1}\approx n_p\sigma_{pp}c \kappa_{pp},
\end{equation}
where $\sigma_{pp}$ and $\kappa_{pp}$ are the cross section and inelasticity for $pp$ interactions, respectively. $\sigma_{pp}$ was given in Ref. \cite{2014PhRvD..90l3014K}, and $\kappa_{pp}$ is set to be 0.5. The photomeson production rate is 
\begin{equation}
t_{p\gamma}^{-1}=\frac{c}{2\gamma_p^2}\int_{\overline{\varepsilon}_{\rm th}}^\infty{d}\overline{\varepsilon}_\gamma\sigma_{p\gamma}\kappa_{p\gamma}\overline{\varepsilon}_\gamma\int_{\overline{\varepsilon}_\gamma/(2\gamma_p)}^\infty{d}\varepsilon_\gamma \varepsilon_\gamma^{-2}\frac{dn_\gamma}{d\varepsilon_\gamma},\label{eq:tpg}
\end{equation}
where $\gamma_p=\varepsilon_p/(m_pc^2)$, $\overline \varepsilon_{p,\rm{th}}\simeq145$~MeV is the threshold energy for the photomeson production, $\overline \varepsilon_\gamma$ is the photon energy in the proton rest frame, and $\sigma_{p\gamma}$ and $\kappa_{p\gamma}$ are the cross section and inelasticity for photomeson production, respectively. We use fitting formulas based on GEANT4 for $\sigma_{p\gamma}$ and $\kappa_{p\gamma}$ (see Ref.~\cite{MN06b}).  The Bethe-Heitler cooling rate is also estimated by Equation (\ref{eq:tpg}) using $\sigma_{\rm BH}$ and $\kappa_{\rm BH}$ instead of $\sigma_{p\gamma}$ and $\kappa_{p\gamma}$, respectively. We use the fitting formulas given in Refs. \cite{SG83a} and \cite{CZS92a} for $\sigma_{\rm BH}$ and $\kappa_{\rm BH}$, respectively.  The synchrotron cooling rate is estimated to be
\begin{equation}
t_{\rm syn}^{-1}=\frac{\gamma_p \sigma_T B^2}{6\pi m_p c}\left(\frac{m_e}{m_p}\right)^2.
\end{equation}
The total cooling rate is given by the sum of all the cooling rates.

\begin{figure*}[tb]
   \begin{center}
    \includegraphics[width=\linewidth]{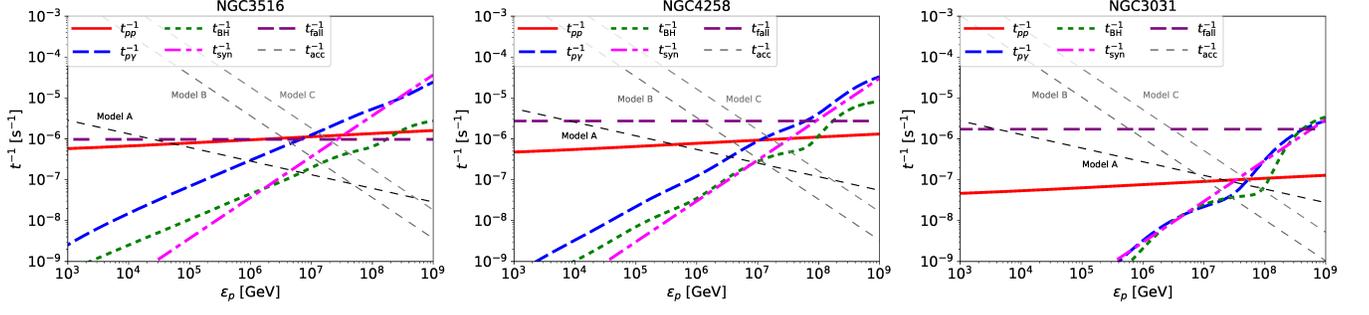}
    \caption{The cooling, escape, and acceleration rates for NGC 3516 (left), NGC 4258 (middle), and NGC 3031 (right).}
    \label{fig:time}
   \end{center}
  \end{figure*}

Figure \ref{fig:time} shows the loss and acceleration rates as a function of proton energy for NGC 3516, NGC 4258, and NGC3031, which have $\dot m\sim 0.9\dot m_{\rm cr}$,  $\dot m\sim 0.3\dot m_{\rm cr}$, and $\dot m\sim 0.04\dot m_{\rm cr}$ respectively. For NGC 3516, $t_{\rm fall}$ and $t_{pp}$ are comparable in the entire energy range. The photomeson production is effective above $\varepsilon_p\gtrsim 30$ PeV. The synchrotron and Bethe-Heitler losses are always subdominant in the range of our interest. On the other hand, for NGC 4258 and NGC 3031, the infall timescale is always dominant below the cutoff energy due to lower $\dot m$. Note that the critical energy at which $t_{\rm acc}=t_{\rm loss}$ is very low for model A, compare to the other models. Such a lower critical energy is required to achieve a cutoff energy similar to the other models (see Figure \ref{fig:ELE}) because the stochastic acceleration results in a hard spectrum with a gradual cutoff (cf. Refs.~\cite{bld06,kmt15}).

To understand the parameter dependences of each timescale, we write $t_{\rm p\gamma}^{-1}\sim n_{\varepsilon_\gamma} \kappa_{p\gamma}\sigma_{p\gamma}c$, where $n_{\varepsilon_\gamma}\approx L_{\varepsilon_\gamma}/(2\pi R^2 c \varepsilon_\gamma)$ is the differential photon number density and $L_{\varepsilon_\gamma}$ is the differential photon luminosity. Then, if we fix the parameters in Table \ref{tab:fixed}, the parameter dependence of the loss rates are $t_{\rm fall}^{-1}\propto M_{\rm BH}^{-1}$, $t_{pp}^{-1}\propto \dot m M_{\rm BH}^{-1}$, $t_{p\gamma}^{-1}\propto \dot m^2 M_{\rm BH}^{-1}$, and $t_{\rm syn}^{-1}\propto\dot m M_{\rm BH}^{-1}$. Interestingly, all the loss rates are proportional to $M_{\rm BH}^{-1}$, while they have a different $\dot m$ dependence. For the case with $\dot m \sim \dot m_{\rm crit}$ as in NGC 3516, $t_{p\gamma}^{-1} \lesssim t_{pp}^{-1}$ and $t_{pp}^{-1}\sim t_{\rm fall}$ below the cutoff energy. Since a lower value of $\dot m$ makes $t_{\rm fall}$ shorter and $t_{p\gamma}$ longer relative to $t_{pp}$, we can approximately use $t_{\rm fall}$ as the energy loss timescale, and $pp$ collisions are the main channel of neutrino production for $\dot m \lesssim \dot m_{\rm crit}$. We describe analytic estimates with this approximation in Section \ref{sec:neutrino}.

Figure \ref{fig:ELE} shows the resulting proton spectrum, $E_pF_{E_p}=\varepsilon_p L_{\varepsilon_p}/(4\pi d_L^2)$, and the injection proton spectrum, $E_pF_{E_p,\rm inj}=\varepsilon_p^2 \dot N_{\varepsilon_p,\rm inj}/(4\pi d_L^2)$, where $E_p$ is the energy in the observer's frame. Since we focus on the very nearby objects, we ignore the effect of redshift, i.e., $E_p\approx \varepsilon_p$.   The parameter sets are tabulated in Tables \ref{tab:param} and \ref{tab:fixed}. We choose these parameter sets so that our model can reproduce the diffuse MeV gamma-ray and TeV--PeV neutrino intensities (see the accompanying paper). We also tabulate the total proton luminosity, $L_p=\int L_{\varepsilon_p} d\varepsilon_p$, and pressure ratio of the non-thermal to thermal components, $P_{\rm CR}/P_g = \int \varepsilon_p N_{\varepsilon_p}d\varepsilon_p /(6\pi R^2 H m_pn_p C_s^2)$. To achieve the observed diffuse neutrino intensity, we need $P_{\rm CR}/P_g\sim 0.1$ for models A and B,  while $P_{\rm CR}/P_g\sim 0.5$ for model C.

In model A, the stochastic acceleration model leads to a hard spectrum below the critical energy, which is $E_pF_{E_p}\propto \varepsilon_p^{3-q}$. Above the critical energy, the spectrum gradually becomes softer. For NGC 3516, the photomeson production is efficient above $\varepsilon_p\simeq10^6$ GeV, which makes a sharp cutoff. For NGC 4258 and NGC 3031, the cooling processes are inefficient. This leads to a more gradual cutoff, resulting in a higher peak energy than that for NGC 3516. 
In models B and C, the resulting spectra are very similar to the injection spectra, because the infall is the dominant loss process. In this case, the proton number spectrum in the RIAF is written as $N_{\varepsilon_p}\approx \dot N_{\varepsilon_p,\rm inj}t_{\rm fall}$, leading to $L_{\varepsilon_p}\approx \varepsilon_p \dot N_{\varepsilon_p,\rm inj}$. For NGC 3516, we can see a slight difference between the two spectra due to the $pp$ cooling. Note that we cannot observe this flux of protons on Earth because of the energy loss processes and deflection by interstellar and intergalactic magnetic fields.

\section{High-energy Neutrinos}\label{sec:neutrino}

\subsection{Meson cooling}

We numerically calculate the neutrino production through both photomeson and hadronuclear interactions. The neutrinos are produced by decay of pions and muons. In general the high-energy neutrinos can be suppressed by meson and muon cooling, when their lifetimes are longer than the cooling time. Here, we estimate the hadronic cooling time for pions and synchrotron cooling for pions and muons. The hadronic cooling rate for pions is estimated to be $t_{\pi p}^{-1}\sim n_p \sigma_{\pi p}\kappa_{\pi p}c$, where $\sigma_{\pi p}\sim 50$ mb and $\kappa_{\pi p}\sim0.8$ are the pion-proton interaction cross section and inelasticity, respectively. The critical energy above which the pion hadronic cooling is efficient is $\varepsilon_{\nu,\rm \pi p}\approx m_\pi c^2/(n_p\sigma_{\pi p}\kappa_{\pi p}c\tau_{\pi0}) \sim 2\times10^{21} \mathcal R_1^{3/2}\alpha_{-1}M_8\dot m_{-2}^{-1}$ eV, where $m_\pi$ and $\tau_{\pi0}$ are the mass and  decay time of pions, respectively. Thus, we can safely ignore the pion hadronic cooling.

The synchrotron cooling time for a particle $i$ is written as $t_{i,\rm syn}\approx 6\pi m_i^5 c^5/(m_e^2\sigma_Tc\varepsilon_i^2B^2)$, where $m_i$ and $\varepsilon_i$ are the mass and energy of the particle. Equating the lifetime and synchrotron cooling time, we can estimate the critical energies above which the synchrotron cooling is effective to be $\varepsilon_{\nu,\pi\rm syn}=\sqrt{3\pi m_\pi^5c^5/(8m_e^2\sigma_TB^2\tau_\pi)}\simeq 1.0\times10^{17} \mathcal R_1^{5/4}\alpha_{-1}^{1/2}\beta_{0.3}^{1/2}M_8^{1/2}\dot m_{-2}^{-1/2}$ eV for pions and $\varepsilon_{\nu,\mu\rm syn}=\sqrt{2\pi m_\mu^5c^5/(3m_e^2\sigma_TB^2\tau_\mu)}\simeq 7.5\times10^{15} \mathcal R_1^{5/4}\alpha_{-1}^{1/2}\beta_{0.5}^{1/2}M_8^{1/2}\dot m_{-2}^{-1/2}$ eV for muons.
Here, $m_\mu$ and $\tau_{\mu0}$ are the mass and decay time of muons, respectively. Since we are interested in TeV -- PeV neutrinos, we will ignore the cooling effect by mesons and muons. 

\subsection{Neutrino spectrum}

To calculate high-energy neutrino spectra from $pp$ interactions, we use the method given by Ref.~\cite{kab06}, where the $pp$-neutrino spectrum, $L_{pp,\varepsilon_\nu}= \varepsilon_\nu t_{\rm pp}^{-1} dN/d\varepsilon_\nu$, is given by 
\begin{equation}
 \frac{L_{pp,\varepsilon_\nu}}{\varepsilon_\nu} \approx c n_p \int_{\varepsilon_\nu}^{\infty}\sigma_{pp}(\varepsilon_p)N_{\varepsilon_p}\mathcal H_\nu\left(\frac{\varepsilon_\nu}{\varepsilon_p},~\varepsilon_p\right)\frac{d\varepsilon_p}{\varepsilon_p},
\end{equation}
where $\mathcal H_\nu(\varepsilon_\nu/\varepsilon_p,~\varepsilon_p)$ is the spectral shape of the neutrinos from mono-energetic protons of $\varepsilon_p$ (see Ref. \cite{kab06} for details). This method is valid only for $\varepsilon_\nu > 100$ GeV. Since our scope is to discuss the detection prospects by IceCube-like detectors, we focus on neutrinos above 100 GeV. For $p\gamma$ neutrinos, we approximately calculate the spectrum using the semi-analytic formalism of Refs. \cite{KMM17b,KMB18a}, including the physical processes described in the previous section. Ignoring the effects of the meson cooling, the $p\gamma$-neutrino spectrum is given by 
\begin{equation}
 \varepsilon_\nu L_{p\gamma,\varepsilon_\nu} \approx \frac{3}{8}f_{p\gamma} \varepsilon_p L_{\varepsilon_p},
\end{equation}
where $\varepsilon_\nu \approx 0.05\varepsilon_p$ and $f_{p\gamma}\approx t_{p\gamma}^{-1}/t_{\rm loss}^{-1}$.
  The neutrino flavor ratio at the sources is $(\nu_e,~\nu_\mu,~\nu_\tau)=(1,~2,~0)$ owing to the inefficient muon and pion cooling.  The neutrinos change their flavors to $(\nu_e,~\nu_\mu,~\nu_\tau)=(1,~1,~1)$ during the propagation to the Earth through neutrino oscillation, and thus, the muon neutrino flux is a factor of 3 lower than the total neutrino flux. 

Figure \ref{fig:ELE} shows the resulting muon neutrino fluxes, 
\begin{equation}
E_{\nu_\mu} F_{E_{\nu_\mu}}\approx\frac{\varepsilon_\nu L_{\varepsilon_\nu}}{12\pi d_L^2 },\label{eq:EFEnu}
\end{equation}
where $L_{\varepsilon_\nu}=L_{pp,\varepsilon_\nu}+L_{p\gamma,\varepsilon_\nu}$. Since the $pp$ neutrino decay spectrum is softer than the parent proton spectrum for models A and B, these two models give similar neutrino spectral shapes. The neutrinos produced by $pp$ interaction are dominant for the low energy range, but the photomeson production gives a comparable contribution around the cutoff energy for the cases with $\dot m \gtrsim 0.01$ (NGC 3516 and NGC 4258). For NGC 3031, $\dot m$ is too low to effectively create neutrinos via photomeson production.

\subsection{Analytic estimate}\label{sec:analytic}

We can approximately derive analytic estimates of the neutrino flux from LLAGNs for the power-law injection cases. When infall is the dominant loss process, we can write $N_{\varepsilon_p}\approx t_{\rm fall} \dot N_{p,\rm inj}$, as discussed in the previous section. Then, the proton luminosity is approximated to be 
\begin{equation}
 \varepsilon_p L_{\varepsilon_p} \approx \varepsilon_p^2 \dot N_{p,\rm inj},\label{eq:ELEnu}
\end{equation}
and the normalization is determined by $\int L_{\varepsilon_p} d\varepsilon_p= \epsilon_p\dot m L_{\rm Edd}\propto \epsilon_p \dot m M_{\rm BH}$.  The neutrino production efficiency is given by 
\begin{equation}
f_{pp}\approx {\rm min}(1~,\frac{t_{\rm fall}}{t_{pp}})\approx \frac{8\dot m}{\alpha^2}\frac{\sigma_{pp}\kappa_{pp}}{\sigma_T}\simeq 0.36\alpha_{-1}^{-2}\dot m_{-2},\label{eq:fpp}
\end{equation}
where we use $\sigma_{pp}\sim 60$ mb and $\kappa_{pp}\sim0.5$ for the estimate, which corresponds to the values for $\varepsilon_p\sim 1-10$ PeV. $f_{pp}$ becomes unity around the saturation accretion rate,
\begin{equation}
\dot m_{\rm sat}\sim 2.8\times10^{-2}\alpha_{-1}.
\end{equation}
With our reference parameters, this accretion rate is very close to the critical accretion rate, $\dot m_{\rm crit}$.
The all-flavor differential neutrino luminosity is approximated to be 
\begin{eqnarray}
\varepsilon_\nu L_{\varepsilon_\nu} \approx \frac12 f_{pp}\varepsilon_p L_{\varepsilon_p}\propto \epsilon_p \dot m^2 \alpha^{-2}M_{\rm BH}\propto L_X \epsilon_p,\label{eq:ELEnu2}
\end{eqnarray}
where $\varepsilon_\nu\approx0.04\varepsilon_p$.
Interestingly, the neutrino luminosity is proportional to $L_X$ and $\epsilon_p$, and independent of the other parameters. The differential muon neutrino energy flux is computed using Equations (\ref{eq:dotNinj}), (\ref{eq:EFEnu}), (\ref{eq:ELEnu}), (\ref{eq:fpp}), and (\ref{eq:ELEnu2}).
This method approximates the peak $pp$-neutrino flux within an error of factors of 2 and 1.3 for $s_{\rm inj}=1.0$ and 2.0, respectively.

\subsection{Detectability of neutrinos from nearby LLAGNs}\label{sec:detect}

We evaluate the number of through-going muon track events following Refs. \cite{Laha:2013eev,Murase:2016gly}. We estimate the differential detection rate of through-going tracks: 
\begin{equation}
 \frac{d\dot {\mathcal N_\mu}}{dE_\mu}\approx \frac{{\mathcal N}_A\mathcal A_{\rm det}}{\alpha_\mu+\beta_\mu E_\mu}\int_{E_\mu}^\infty dE_\nu\phi_{\nu_\mu}\sigma_{\rm CC}e^{-\tau_{\nu N}},
\end{equation}
where $E_\nu$ is the incoming neutrino energy, $E_\mu$ is the muon energy, ${\mathcal N}_A$ is the Avogadro number, $\mathcal A_{\rm det}$ is the muon effective area, $\sigma_{\rm CC}$ is the charged-current cross section, $\tau_{\nu N}$ is the optical depth to neutrino-nucleon scatterings in the Earth, and the denominator in the right-hand side indicates the muon energy loss rate (see Ref. \cite{Murase:2016gly} and references therein).  This method can reproduce the effective area reported by Ref. \cite{Aartsen:2014cva}. We evaluate the background including both the conventional and the prompt atmospheric muon neutrinos.

\begin{figure*}[tb]
   \begin{center}
    \includegraphics[width=\linewidth]{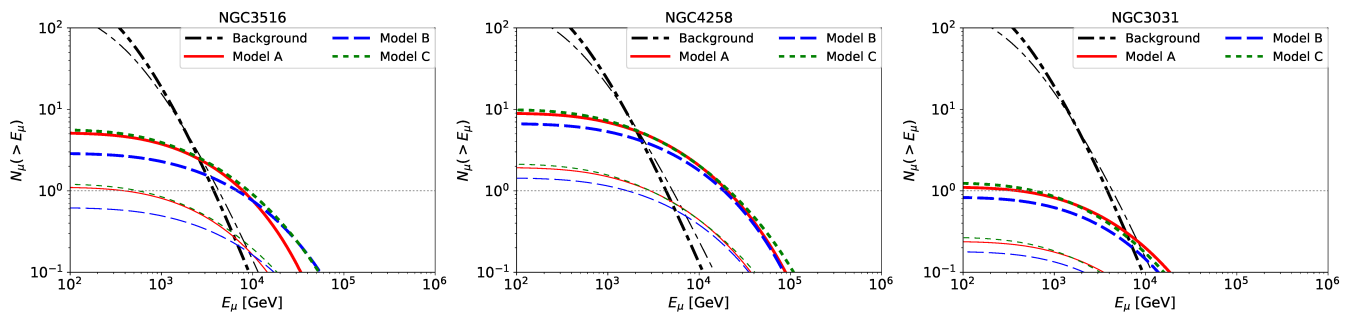}
    \caption{The expected number of through-going track events from NGC 3516  (left panel), NGC 4258 (middle panel), and NGC 3031 (right panel) for models A (solid), B (dashed), and C (dotted) for a 10-year operation of IceCube-Gen2 (thick lines) and for IceCube (thin lines). The dot-dashed lines show the expected background.}
    \label{fig:Nmu}
   \end{center}
  \end{figure*}

We plot ${\mathcal N}_\mu (>E_\mu) = \int_{E_\mu}^{\infty} dE_\mu' \int dt d\dot {\mathcal N_\mu}/dE_\mu'$ in Figure \ref{fig:Nmu} for a ten-year operation with IceCube and IceCube-Gen2 for NGC 3516, NGC 4258, and NGC 3031. IceCube cannot detect signals from individual objects due to lower effective area. IceCube-Gen2 can detect the signals from NGC 4258, while it is challenging to detect NGC 3516. Although NGC 3516 has a neutrino flux comparable to that of NGC 4258, the higher declination causes the lower $N_\mu(>E_\mu)$ due to the Earth attenuation, especially in Model B. The neutrino emission from NGC 3031 is too faint to be detected even with IceCube-Gen2.

\begin{figure}[tb]
   \begin{center}
    \includegraphics[width=\linewidth]{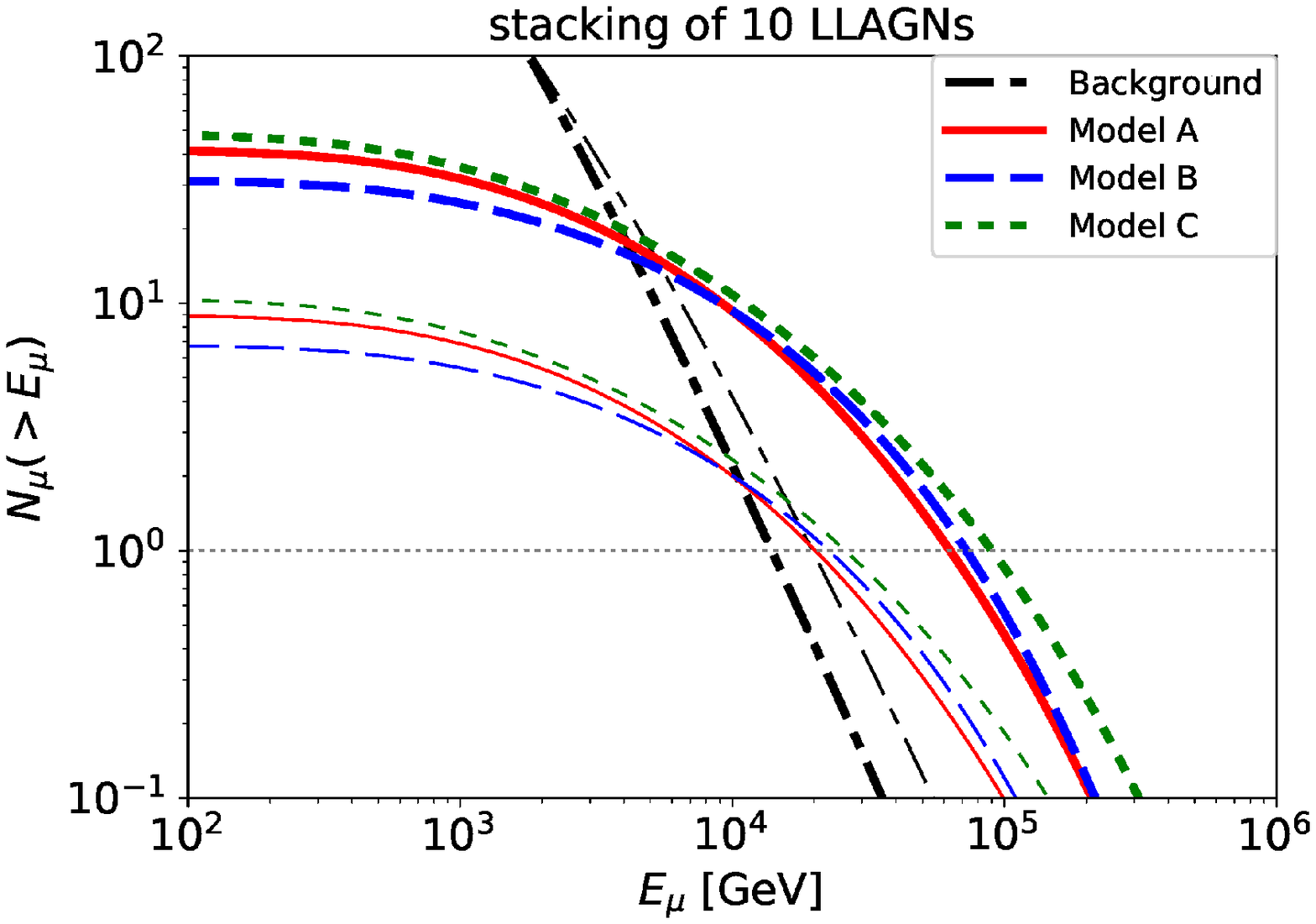}
    \includegraphics[width=\linewidth]{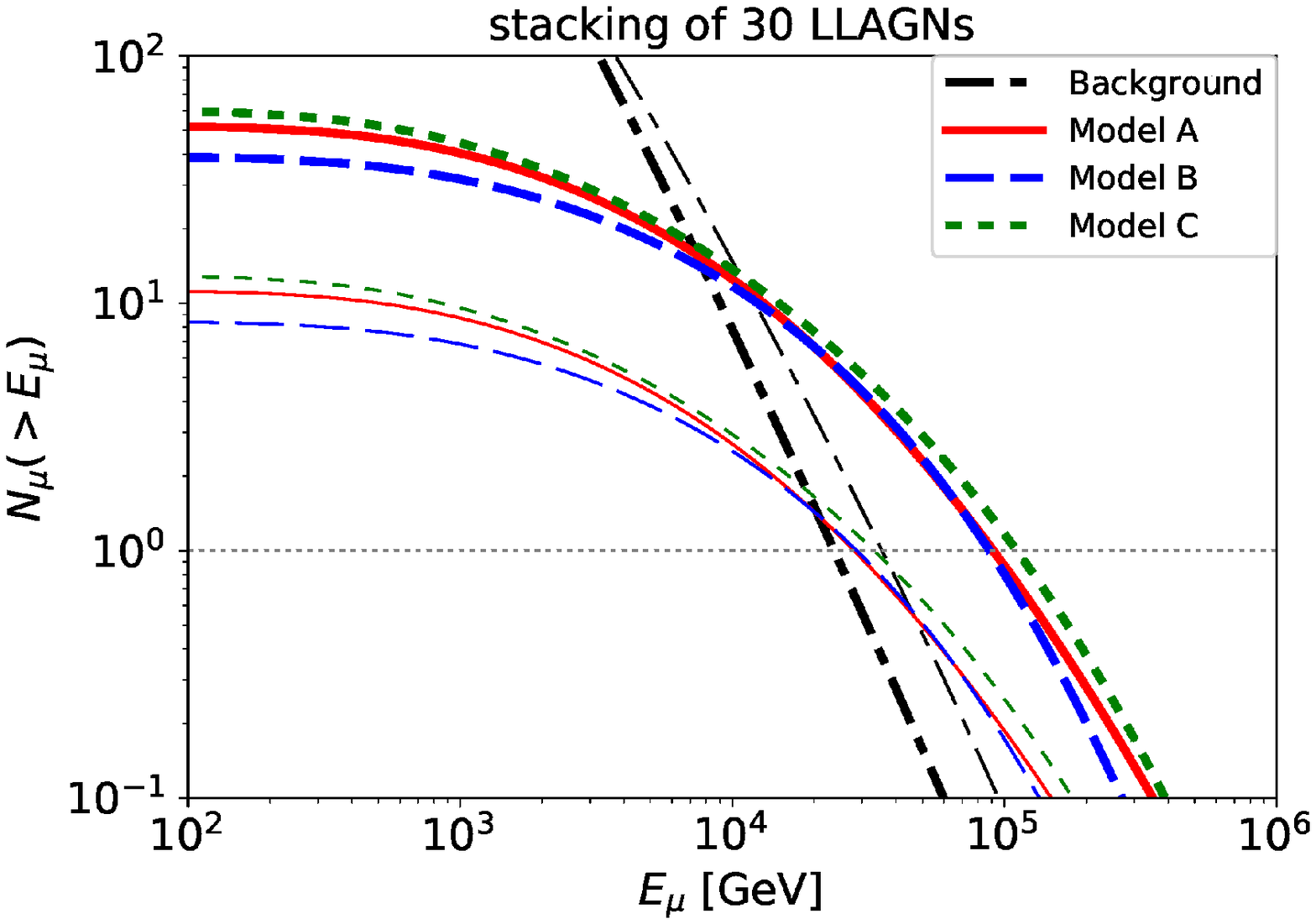}
    \caption{Same as Figure \ref{fig:Nmu}, but stacking 10 (upper panel) and 30 LLAGNs (lower panel).}
    \label{fig:stack}
   \end{center}
  \end{figure}

Since the neutrino flux is roughly proportional to the X-ray flux, we place the LLAGNs listed in Ref. \cite{2018A&A...616A.152S} in order of the X-ray flux, as shown in Table \ref{tab:param}, and estimate the number of track events above $E_\mu$ by stacking them.  Figure \ref{fig:stack} shows the resulting event number for a 10-year operation with IceCube-Gen2 and IceCube by stacking 10 LLAGNs and 30 LLAGNs. With IceCube-Gen2, we expect 3 -- 7 events above 30 TeV where the background is negligible. 
Interestingly, the neutrinos from the ten brightest LLAGNs will be sufficient for the detection, because stacking more LLAGNs leads to an increase of the atmospheric background. With the current IceCube experiment, the effective area and angular resolution are $10^{2/3}$ times smaller and 3 -- 5 times larger than those of IceCube-Gen2, respectively. Then, the event number is about 4 -- 5 times lower and the background rate is 10 -- 30 times higher, making the detection of neutrinos more challenging, as seen in the figures.

\section{Cascade gamma-ray emission}\label{sec:gamma}

Hadronuclear and photohadronic processes produce very-high-energy (VHE) gamma rays through neutral pion decay and high-energy electron/positron pairs through charged pion decay and the Bethe-Heitler process. The VHE gamma rays are absorbed by soft photons through the $\gamma\gamma\rightarrow e^+ e^-$ process in the RIAF, and produce additional high-energy electron/positron pairs. The high-energy $e^+e^-$ pairs also emit gamma-rays through synchrotron, inverse Compton scattering, and bremsstrahlung, leading to electromagnetic cascades. We calculate the cascade emission by solving the kinetic equations of photons and electron/positron pairs (see Refs.~\cite{2018PhRvD..97h1301M,2019ApJ...874...80M,2019arXiv190404226M}): 
\begin{eqnarray}
 \frac{\partial n^e_{\varepsilon_e}}{\partial t} +\frac{\partial}{\partial \varepsilon_e}\left[\left(P_{\rm IC}+P_{\rm syn}+P_{\rm ff}+P_{\rm Cou}\right)n^e_{\varepsilon_e}\right]\nonumber\\
 =\dot n_{\varepsilon_e}^{(\gamma\gamma)}-\frac{n^e_{\varepsilon_e}}{t_{\rm esc}} +\dot n_{\varepsilon_e}^{\rm inj},
\end{eqnarray}
\begin{equation}
 \frac{\partial n^\gamma_{\varepsilon_\gamma}}{\partial t} = -\frac{n_{\varepsilon_\gamma}^\gamma}{t_{\gamma\gamma}}  - \frac{n_{\varepsilon_\gamma}^\gamma}{t_{\rm esc}} + \dot n_{\varepsilon_\gamma}^{(\rm IC)}+ \dot n_{\varepsilon_\gamma}^{(\rm ff)}+ \dot n_{\varepsilon_\gamma}^{(\rm syn)} + \dot n_{\varepsilon_\gamma}^{\rm inj},
\end{equation}
where $n_{\varepsilon_i}^i$ is the differential number density ($i=e$ or $\gamma$), $\dot n_{\varepsilon_i}^{(xx)}$ is the particle source term from the process $xx$  ($xx=\rm IC$ (inverse Compton scattering), $\gamma\gamma$ ($\gamma\gamma$ pair production), syn (synchrotron), or ff (bremsstrahlung)), $\dot N_{\varepsilon_i}^{\rm inj}$ is the injection term from the hadronic interaction, and $P_{yy}$ is the energy loss rate for the electrons from the process $yy$ ($yy=\rm IC$ (inverse Compton scattering), syn (synchrotron), ff (bremsstrahlung), or Cou (Coulomb collision)). We calculate the cascade spectra using spherical coordinates, while the other calculations are made in cylindrical coordinates. The effect of geometry has little influence on our results. 

Here, we approximately treat the injection terms of photons and pairs from hadronic interactions. The injection terms for photons and pairs consist of the sum of the relevant processes: $\dot n_{\varepsilon_\gamma}^{\rm inj}=\dot n_{\varepsilon_\gamma}^{(p\gamma)}+\dot n_{\varepsilon_\gamma}^{(pp)}$ and $\dot n_{\varepsilon_e}^{\rm inj}=\dot n_{\varepsilon_e}^{({\rm BH})}+\dot n_{\varepsilon_e}^{(p\gamma)}+\dot n_{\varepsilon_e}^{(pp)}$. We approximate the terms due to Bethe-Heitler and $p\gamma$ processes to be
\begin{eqnarray}
 \varepsilon_\gamma^2\dot n_{\varepsilon_\gamma}^{(p\gamma)} \approx \frac12 t_{p\gamma}^{-1}\varepsilon_p^2n_{\varepsilon_p},\\
 \varepsilon_e^2\dot n_{\varepsilon_e}^{(p\gamma)} \approx \varepsilon_\nu^2 n_{\varepsilon_\nu}^{(p\gamma)} \approx\frac18 t_{\rm p\gamma}^{-1} \varepsilon_p^2 n_{\varepsilon_p}, \\
 \varepsilon_e^2\dot n_{\varepsilon_e}^{({\rm BH})} \approx t_{\rm BH}^{-1} \varepsilon_p^2 n_{\varepsilon_p},
\end{eqnarray}
where $\varepsilon_\gamma\approx 0.1 \varepsilon_p$ and $\varepsilon_e\approx 0.05\varepsilon_p$ for photomeson production, and $\varepsilon_e\approx (m_e/m_p)\varepsilon_p$ for Bethe-Heitler process. For the injection terms from $pp$ interactions, see Ref. \cite{2019ApJ...874...80M}.

We plot proton-induced cascade gamma-ray spectra in Figure \ref{fig:ELE}. A sufficiently developed cascade emission generates a flat spectrum below the critical energy at which $\gamma\gamma$ attenuation becomes ineffective. The optical depth to the electron-positron pair production is estimated to be 
\begin{equation}
\tau_{\gamma\gamma}(\varepsilon_\gamma) \approx R \int \mathcal K(x) \frac{dn_\gamma}{d\varepsilon_\gamma} d\varepsilon_\gamma,
\end{equation}
where $\varepsilon_\gamma$ is the gamma-ray energy, $\mathcal K(x)=0.652 \sigma_T (x-x^{-2}) \ln(x) H(x-1)$, $x=\varepsilon_\gamma \mathcal \varepsilon_\gamma/(m_ec^2)$, and $H(x)$ is the Heaviside step function \cite{cb90}.  We tabulate the values of the critical energy, $\varepsilon_{\gamma\gamma}$, at which $\tau_{\gamma\gamma}=1$ in Table \ref{tab:quant}. We can see flat spectra below the critical energy. Note that the tabulated values are approximately calculated using a fitting formula, while the cascade calculations are performed with the exact cross section.  We overplot the {\it Fermi} LAT sensitivity curve in the high galactic latitude region with a 10-year exposure obtained from Ref. \cite{2017ExA....44...25D}. The predicted fluxes are lower than the sensitivity curve for all the cases. The Cherenkov Telescope Array (CTA) has a better sensitivity above 30 GeV than LAT, but the cascade gamma-ray flux is considerably suppressed in the VHE range due to the $\gamma\gamma$ attenuation. For a lower $\dot m$ object that has a higher value of $\varepsilon_{\gamma\gamma}$, such as NGC 5866, the cascade flux is too low to be detected by CTA. Therefore, it would be challenging to detect the cascade gamma rays with current and near-future instruments,  except for Sgr A*. 

Sgr A* has two distinct emission phases: the quiescent and flaring states (see Ref. \cite{2010RvMP...82.3121G} for review). The X-ray emission from the quiescent state of Sgr A* is spatially extended to $\sim$ 1'', which corresponds to $10^5R_S$ for a black hole of $4\times10^6 \rm~M_{\odot}$ \cite{2003ApJ...591..891B}. Hence, our model is not applicable to the quiescent state. On the other hand, the flaring state of Sgr A* shows $10-300$ times higher flux than the quiescent state with the time variability of $\sim 1$ h \cite{2003A&A...407L..17P}. This variability timescale implies that the emission region should be $\lesssim10^2R_S$. However, the value of $\dot m$ for the brightest flare estimated by Equation (\ref{eq:mdot}) is less than $10^{-4}$. Since our model is not applicable to such a low-accretion-rate system (see Section \ref{sec:RIAF}), we avoid discussing it in detail. The detailed estimate should be made in the future (see Ref. \cite{2019ApJ...879....6R} for related discussion).

\section{Summary}\label{sec:summary}

We have investigated high-energy multi-messenger emissions, including the MeV gamma-rays, high-energy gamma-rays, and neutrinos, from nearby individual LLAGNs, focusing on their multi-messenger detection prospects. 
We have refined the RIAF model of LLAGNs, referring to recent simulation results. Our one-zone model is roughly consistent with the observed X-ray features, such as an anti-correlation between the Eddington ratio and the spectral index. RIAFs with $\dot m \gtrsim0.01$ emit strong MeV gamma rays through Comptonization, which will be detected by the future MeV satellites such as e-ASTROGAM, AMEGO, and GRAMS.

We have also calculated the neutrino and cascade gamma-ray spectra from accelerated protons. We considered three models for the proton spectrum. In model A, we considered stochastic acceleration by turbulence and solve the diffusion equation in momentum space. In models B and C, we do not specify the acceleration mechanism and assumed an injection term with a power-law and an exponential cutoff. Using such proton spectra, we have numerically calculated the neutrino spectra, taking account of the relevant cooling processes and the decay spectra. 
Since $pp$ inelastic collisions provide the main channel for high-energy neutrino production, the neutrino spectrum follows the proton spectrum. Close to the cutoff energy, $\varepsilon_\nu\sim100$ TeV, the photomeson production is as efficient as $pp$ interactions, leading to a  comparable contribution to the neutrino flux. 
With a few to 10 LLAGNs stacked, a 10-year operation of IceCube-Gen2 will enable us to detect a few to several neutrinos from LLAGNs, otherwise they will put meaningful constraints on the parameter space. On the other hand, the cascade emission is difficult to detect with {\it Fermi} or CTA. Bright objects have a lower $\gamma\gamma$ cutoff energy, while objects with a higher value of the cutoff energy are too dim to produce a detectable signal. 

AGN coronae and RIAFs are thought to be promising sites of particle acceleration, and accompanying papers suggest the AGN cores as the main origin of the mysterious 10 -- 100~TeV component in the diffuse neutrino flux observed in IceCube~\cite{2019arXiv190404226M}. The model predicts that both Seyfert galaxies and LLAGNs are promising sources of high-energy neutrinos and MeV gamma rays. Our studies suggest the relevance of multi-messenger searches for LLAGNs whether the 10 -- 100~TeV neutrinos mainly come from Seyfert galaxies or LLAGNs.


\medskip
\begin{acknowledgments}
This work is supported in part by  JSPS Oversea Research Fellowship, JSPS Postdoctral Fellowship, the IGC postdoctoral fellowship program (S.S.K.), the Alfred P. Sloan Foundation, NSF Grant No. PHY-1620777 (K.M.), and the Eberly Foundation (P.M.).
\end{acknowledgments}


\bibliography{ssk}
\bibliographystyle{apsrev4-1}
\end{document}